# Biomechanical monitoring and machine learning for the detection of lying postures

Silvia Caggiari [a,*], Peter R. Worsley [a], Yohan Payan [b], Marek Bucki [c], Dan L. Bader [a]

[a] Skin Health Research Group, School of Health Sciences, Faculty of Environmental and Life Sciences, University of Southampton, SO17 1BJ, United Kingdom

[b] University of Grenoble Alpes, CNRS, TIMC-IMAG, 38000 Grenoble, France

[c] TexiSense, France

*Corresponding author: E-mail address: S.Caggiari@soton.ac.uk (S. Caggiari)

Abstract

*Background:* Pressure mapping technology has been adapted to monitor over prolonged periods to evaluate pressure ulcer risk in individuals during extended lying postures. However, temporal pressure distribution signals are not currently used to identify posture or mobility. The present study was designed to examine the potential of an automated approach for the detection of a range of static lying postures and corresponding transitions between postures.

*Methods:* Healthy subjects (n = 19) adopted a range of sagittal and lateral lying postures. Parameters reflecting both the interactions at the support surface and body movements were continuously monitored. Subsequently, the derivative of each signal was examined to identify transitions between postures. Three machine learning algorithms, namely Naïve-Bayes, k-Nearest Neighbors and Support Vector Machine classifiers, were assessed to predict a range of static postures, established with a training model (n = 9) and validated with new input from test data (n = 10).

*Findings:* Results showed that the derivative signals provided a means to detect transitions between postures, with actimetry providing the most distinct signal perturbations. The accuracy in predicting the range of postures from new test data ranged between 82%–100%, 70%–98% and 69%–100% for Naïve-Bayes, k-Nearest Neighbors and Support Vector Machine classifiers, respectively.

*Interpretation:* The present study demonstrated that detection of both static postures and their corresponding transitions was achieved by combining machine learning algorithms with robust parameters from two monitoring systems. This approach has the potential to provide reliable indicators of posture and mobility, to support personalised pressure ulcer prevention strategies.

## 1. Introduction

Systems capable of automatically classifying patterns of movement performed by a human subject e.g. wearable actimetry sensors, are widely used in many clinical and research applications in healthcare through advanced human-machine interfaces (Manini and Sabatini, 2010; Mathie et al., 2003; Mohammed et al., 2016). Recently, their use has been shown to improve the provision of optimal turning critical for PU prevention (Ifedili et al., 2018; Pickham et al., 2018). Nonetheless, there are issues with compliance to body worn sensors, and the information gleamed from actimetry does not correspond to interface pressure measurements (Stinson et al., 2018), which currently represents one of the primary means to assess PU risk.

In recent years, interface pressure measurements systems have been adapted to continuously monitor subject-support surface interactions, with the resulting data being used to indirectly classify a range of postures and movements (Duvall et al., 2019; Kim et al., 2018; Wai et al., 2010; Yousefi et al., 2011). However, the predictive power of these algorithms for early PU risk is largely dependent on the magnitude of the applied pressure in pre-determined areas of the pressure sensing mat (Wai et al., 2010) and are commonly associated with arbitrary thresholds. Only a few studies have combined interface pressure measurements and actimetry signals for classifying postures, none of which were directly focused on pressure ulcers (Zemp et al., 2016).

In a recent publication, the authors have identified a series of robust signals estimated from both continuous pressure mapping and actimetry systems, which can accurately track postures and mobility during different evoked postures (Caggiari et al., 2019). However, the signals in isolation demonstrated limited sensitivity and specificity, therefore a combined signal analysis approach was recommended. These signals resulted in large data sets (Bogie et al., 2008), which would benefit from intelligent data processing. While it is well known that actimetry systems can detect posture and mobility with a high degree of accuracy (Edwardson et al., 2016; Lyden et al., 2016), there is limited evidence that parameters estimated from pressure distribution could act as a surrogate for detecting both postures and corresponding transitions between postures during prolonged lying.

Accordingly, the present study was designed to develop a robust methodology for detecting static postures and transitions between postures, using data acquired for pressure monitoring and actimetry systems. Three conventional machine learning algorithms, namely Naïve-Bayes (NB), k-Nearest Neighbors algorithm (KNN) and Support Vector Machine (SVM) classifiers, each of which have been adopted in previous research studies to classify range of postures (Chi-Chun et al., 2008; Duvall et al., 2019; Foubert et al., 2012) were included in the evaluation.

The accuracy for detecting a range of static postures and their corresponding transitions, namely changes in posture, were assessed using the following objectives:

i. Perform data reduction and feature extraction of the raw actimetry and pressure monitoring signals
ii. Create a methodology for the automatically detection of changes in posture from the set of data and
iii. Apply the machine learning algorithms, cross-validated with leave one-out testing, and evaluate their accuracy in classifying the range of prescribed lying postures.

## 2. Methods

2.1. Participants

The training data set was derived from the previous study evaluating the performance of pressure monitoring and actimetry signals to distinguish postures (Caggiari et al., 2019), which was conducted with institutional ethical approval (Ref: 26379). The data from nine of the healthy participants (5 male and 4 female) were allocated into the *training group*, each of whom were observed to performed a number of minimal postural adjustments during the static postures (Caggiari et al., 2019). Participants were aged between 27 and 36 years (mean = 32 years) with an average height and weight of 1.70 m and 72.0 kg (standard deviation = 0.1 m and 17.0 kg), respectively. The corresponding BMIs ranged between 19 and 30 kg/m$^2$.

A separate cohort of ten healthy participants (4 male and 6 female) were recruited into the *test group*, under the same institutional ethics. Participants were aged between 27 and 56 years (mean = 34 years) with an average height and weight of 1.73 m and 68.9 kg (standard deviation = 0.1 m and 15.7 kg), respectively. The corresponding BMIs ranged between 19 and 28 kg/m$^2$.

Exclusion criteria for both groups included participants with a history of skin conditions, neurological or vascular pathologies that could affect tissue health or those were unable to lie in a supine posture for a period of 2 h. Informed consent was obtained from each participant of both groups prior to testing.

2.2. Test equipment

The equipment and test protocol has been described in the recent paper (Caggiari et al., 2019). To review briefly, interface pressure measurements were recorded using a full body pressure monitoring system (ForeSite PT, XSENSOR Technology Corporation, Canada). The fitted mattress cover incorporates 5664 pressure measuring sensor cells, with a spatial resolution of 15.9 mm, covering a sensing area of 762mm × 1880mm. Each sensor operates within a range of 5-200 mmHg (0.7-26.6 kPa) and an acquisition rate of 1 Hz. Three actimetry sensors (Shimmer Platform, Realtime

Technologies Ltd, Dublin, Ireland) were attached to the sternum and the left and right anterior iliac crests with a Velcro strap. Each device represents a small wireless sensor (53mm × 32mm × 25mm), integrating a tri-axial accelerometer and gyroscope, that records real-time calibrated Euler angles data at 51 Hz (range = 2g).

2.3. Test Protocols

All test procedures were performed in the Biomechanics Testing Laboratory in the Clinical Academic Facility in Southampton General Hospital, where room temperature was maintained at 24º. Participants were requested to wear loose fitting clothing and adopt a series of sagittal and lateral postures on a standard hospital bed frame (Hill-Rom, AvantGuard™) and a castellated foam mattress (Solace Foam Mattress, Invacare UK). A continuous lateral rotational system (CLRS) (Vikta Komfitilt$^R$) placed underneath the support surface enabled left and right 20-25º tilt of the overlying mattress in the lateral plane with an automated 10 min cycle time.

Each of the subjects in the *training group* adopted a series of prescribed sagittal postures held for 10 min, achieved by adjusting the head of the bed (HOB) in 10° increments to a maximum of 60º and then lowering by 10º to supine. In addition, lateral postures were evoked through a continuous lateral rotational system (CLRS). An adapted version of the protocol was used for the *test group* intended to evaluate the performance of the classifiers. Here, sagittal postures were held for 20 min starting in the supine posture followed by raising the HOB angle by 20° increments to a maximum of 60°. The HOB was then lowered in 20° increments to supine. Subsequently lateral postures were adopted through the CLRS system, as for the training group. Interface pressure distribution and actimetry data were continuously recorded throughout the two hours test period for both training and test groups. Participants were instructed to remain as still as possible on the mattress.

2.4. Outcome parameters

The results of the previous study (Caggiari et al., 2019) involving a comprehensive ROC analysis revealed a number of parameters as the most accurate in detecting changes in posture, which included:

- Tilt angles (TA) of the trunk with respect to the sagittal and the lateral planes and
- Percentage variation of contact area (CA) of sensors recording a minimum threshold pressure of 20 mmHg estimated at the whole body ROI.

These parameters were estimated for all subjects from both training and test groups. For the estimation of contact area, pressure readings equal or above a 20 mmHg (2.7kPa) threshold were included, as they were most indicative of evoked postural changes (Caggiari, 2020).

Furthermore, signals from training group were adapted to consider static postures in 20º increments of HOB. As an example, Fig. 1 shows the temporal trend of the variations of contact area and the corresponding pressure distribution during sagittal and lateral postures, for one subject of the training group. It is evident that changes in the HOB angle are reflected in the incremental step changes in the signal and changes in the pressure distribution. In particular, the relative change in signal was dependent on the HOB angle, with angles < 20º revealing reduced variation in contact area. This was also evident for lateral postures. In addition, it is evident that there were differences in the signal dependent on whether the HOB was increasing or decreasing corresponding to 20º HOB and supine postures (Fig. 1).

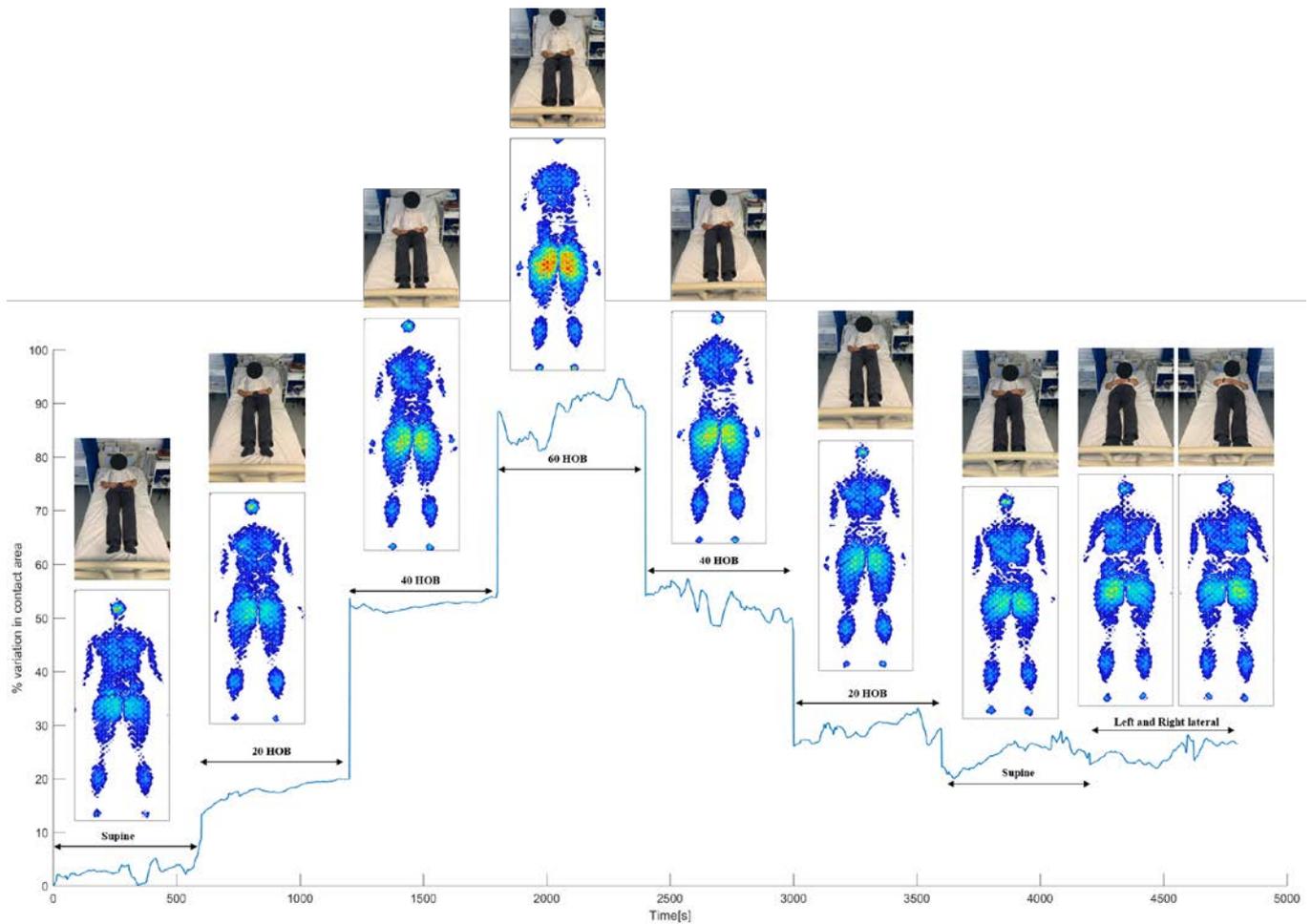

**Fig. 1:** Temporal profile of the percentage variation of contact area at the whole body ROI and the corresponding pressure distribution for 8 sagittal and lateral postures, involving 20º HOB increments.

2.5. Post-processing of the signals from training group

The flowchart in Fig. 2 illustrates the processing of the signals from both training and test groups. This included a moving average filter with a time window of 30 samples for the pressure data from both groups, to remove the high frequency noise. The corresponding actimetry signals, which were originally acquired at 51 Hz, were re-sampled at 1 Hz and filtered using a window of 15 samples (Caggiari, 2019).

The signals from training group were then manually annotated, by denoting the beginning and the end of each evoked posture. The transitions between postures were not included as they contained noise due to natural adjustments in posture observed in all participants. Each of the signals was then interpolated in order to encompass 600 data points, for each posture, resulting in a total of 4800 data points per signal. The interpolation was applied in order to include signals and hence postures of an equivalent time period. All signals were then subjected to a fixed-width sliding window of 60 s to reduce the raw data by evaluating features i.e. mean and derivative values. Mean values were calculated within each sliding window for both trunk tilt angles and contact area signals, resulting in a 80 point data set for each signal. The reduced signals from all the subjects were allocated to the *training data set*. A principal component analysis (PCA) was performed and the training signals projected onto the PCs dimensional space.

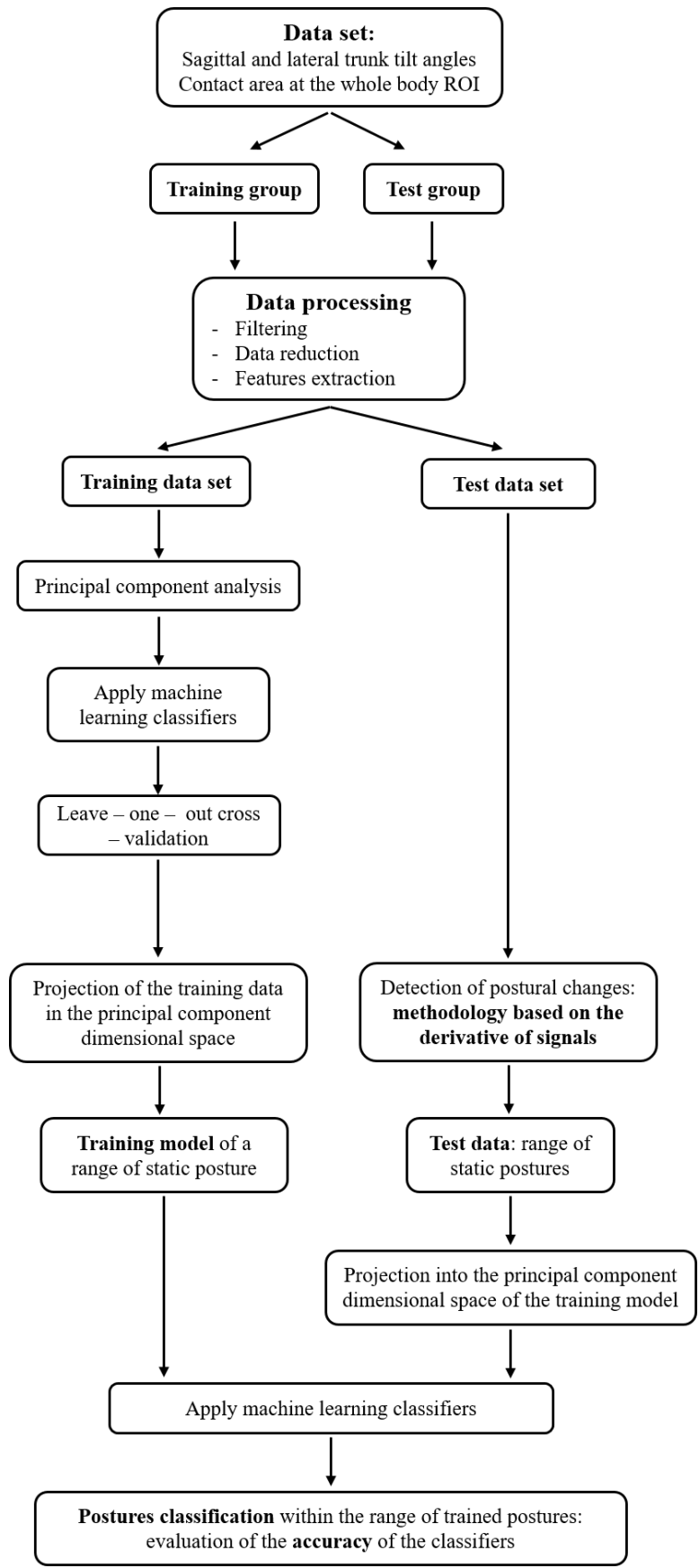

**Fig. 2:** Flow chart depicting the different processes for data acquisition for use with the classifiers.

2.6. Classifiers

Three commonly used classifiers were employed, each of which adopts distinct approaches for event classification.

2.6.1. KNN classifier

The KNN classification rule, first described by Cover and Hart (1967), depends on the distance metric between the new observation point and k nearest data point(s) (Short and Fukunaga, 1981). Given the nature of our data, the Euclidean distance was selected as theGiven the nature of our data, the Euclidean distance was selected as the distance metric. Moreover, the parameter *k* determines how many neighbors will be chosen and its choice has a significant impact on the diagnostic performance of KNN algorithm. Accordingly, a sensitivity analysis was performed and *k* = 10 was identified to provide the highest accuracy and was therefore chosen for the current analysis.

2.6.2. Naïve-Bayes classifier

The Naïve-Bayes classifier represents a probabilistic strategy based on Bayes' theorem, which describes the probability of an event based on the prior knowledge that some other events have already occurred i.e. conditional probability. A Gaussian distribution was considered the most appropriate approach for assessing the conditional probability, which can be written as:

$$P(H \mid E) = \frac{1}{\sqrt{\det(2\pi\sigma_E)}} \exp(-\frac{1}{2}(h-\mu_E)^T \sigma_E^{-1}(h-\mu_E))$$

where H represents a new event and E is some observed event, $\mu_E$ is the mean and $\sigma_E$ is the covariance matrix of the observed events.

2.6.3. Support vector machine (SVM) classifier

The SVM (Burges, 1998) classifier is based on defining a hyperplane that divides the clusters of data. The optimal hyperplane is the one representing the largest distance to the nearest element of each cluster (support vectors). SVM projects the data into a higher dimension from the original space where the hyperplane can be derived from kernel functions. Gaussian kernel function was considered the most appropriate for the present data.

2.6.4. Cross-validation

A leave-one-subject-out cross-validation was performed to test the robustness of the training model in detecting the range of static postures (Fig. 2). This consisted in training a model with data from 8 of the 9 subjects in the training cohort, who were randomly selected. The data of the excluded subject were then tested and the accuracy in postures classification was assessed. This process was repeated for each individual who has been used to test the trained model and the accuracy of the classifiers was determined. Subsequently, a *training model* was created with the set of data of all subjects in the training cohort. The accuracy of all classifiers was then assessed by applying data from subjects in the test cohort. This resulted in a percentage accuracy across all postures adopted in the test data. This percentage accuracy was established for each participant within the test group and calculated as the number of data points correctly classified for each posture with respect to the corresponding total number of points in the signals.

2.7. Test data sets

After filtering, signals from test group were interpolated to encompass 9000 data points (2.5 h of recording), including both static postures and the transition between postures. Signals corresponding to each participant represented a distinct data set to identify changes in posture and test the training model. Each of the test data sets was subjected to the 60-s sliding window for data reduction and both mean and derivative values were estimated within each window prior to identify the changes in posture and classify the corresponding static postures. The derivative signal of both contact area and trunk tilt angles in both sagittal and lateral planes were used for the detection of any change between two postures. The signals were subjected to a discriminant threshold to identify where the variations in the derivative occurred. Different thresholds were examined in order to identify the optimum value which accommodated all subjects. Once the changes in posture from the test data set were identified, the mean signals from the subsequent sliding windows were projected onto the training PCs dimensional space and subjected to posture classification.

# 3. Results

## 3.1. Leave-one-out cross-validation

The cross-validation demonstrated an accuracy ranging between 85%-99% for classification performed with Naïve-Bayes. The corresponding accuracy using KNN and SVM ranged between 53%-95% and 59%-100%, respectively. Accordingly, it is demonstrated that the training data set could provide a robust means for detecting static postures with the test data set.

## 3.2. Detecting changes in posture

Consistent variations in the derivative magnitude of the trunk tilt angles were identified in association with the sagittal changes in posture for all subjects. Smaller variations were also evident in the lateral plane when lateral postures were adopted (data not shown). They were subjected to processing which involved the signal rectification, in order to use a generic positive threshold. The rectified derivative signal for both sagittal and lateral trunk tilt angles were then summarised to obtain a single signal which included both sagittal and lateral changes in posture. An example of the rectified derivative profile of trunk tilt angles for one subject is illustrated in Fig. 3A. The corresponding derivative of the contact area is shown in Fig. 3B. The latter reveals that changes in posture were well distinguished in magnitude at high HOB angles, but less distinctive at lower HOB angles ($< 20°$). Indeed, perturbations in magnitude were also observed during static postures, which did not enable the correct detection of all changes in posture (Fig. 3B). Accordingly, only the derivative of the trunk tilt angles was used and an appropriate discriminant threshold value of 0.10 for all subjects was selected for subsequent detection of the transitions between postures (Fig. 3A), resulting in 100% accuracy for all subjects.

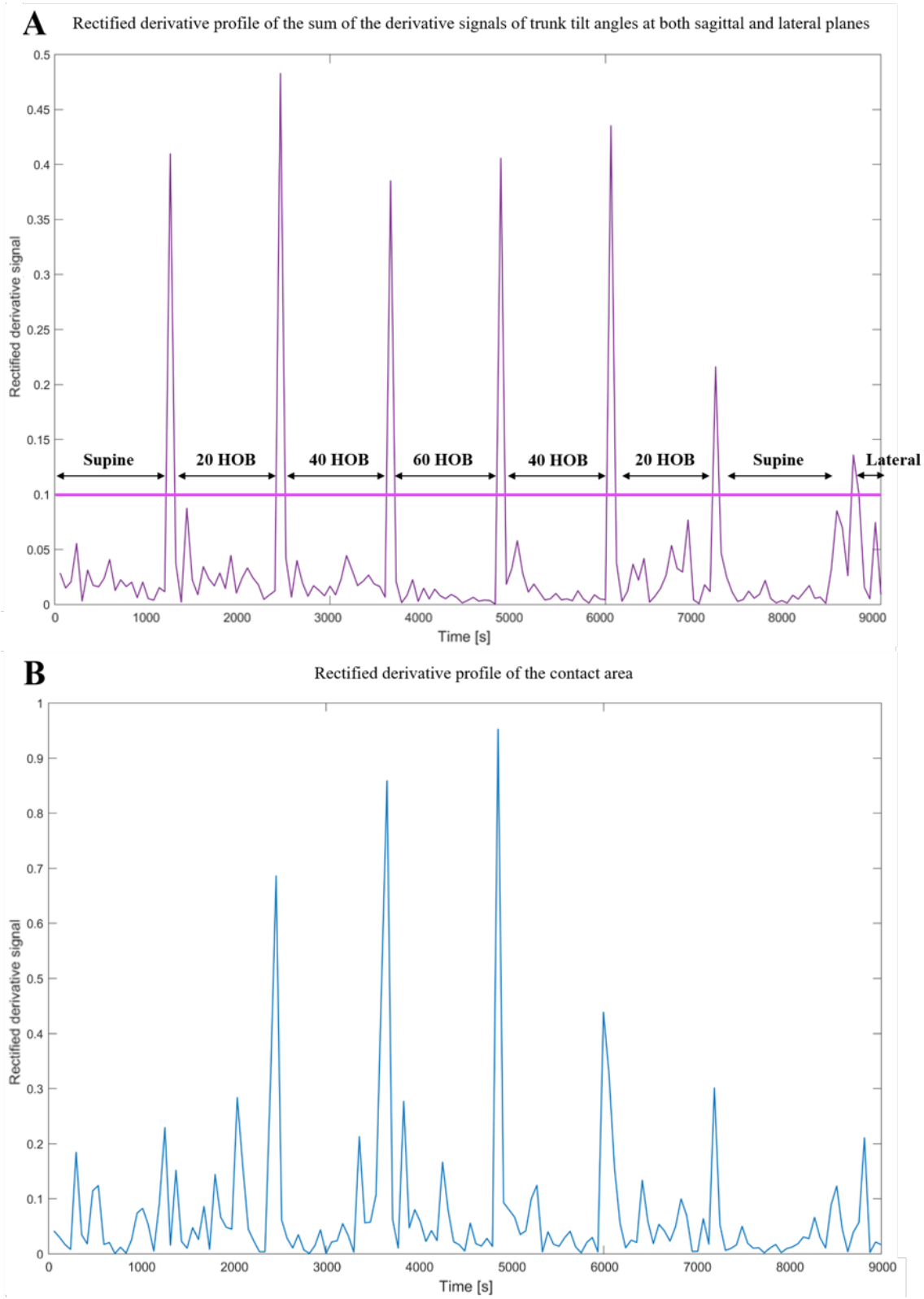

**Fig. 3:** Rectified derivative profile for subject #1 in the test cohort of A) the sum of the derivative signals of trunk tilt angles at both sagittal and lateral planes, B) contact area. Each data point in the signals corresponds to a derivative value calculated within the 60-s sliding window.

3.3. Accuracy in classifying the range of postures

As illustrated in Fig. 4, static postures corresponding to signals for all subjects from the training group (n = 9) resulted in clusters of points spatially distributed across the first and second PCA dimension, which contributed to 87% and 12% of the variance in the signals, respectively. This reflects the changes in signal magnitude associated with either increasing or decreasing the HOB angle (Fig 1). In particular, when a reduced variation in the step changes was observed e.g. HOB < 20º, there is a reduced spatial distribution of adjacent clusters. This is particularly evident in the first principal component (PC1) (x-axis - Fig. 4). When the signals corresponding to the static postures of one subject from the test group are projected onto the training PCs space (data points in pink) separate clusters are observed. These clearly overlapped with the corresponding clusters of points from the training data.

The estimated accuracies for each of the three classifiers are summarised in Table 1. It is evident that for increments of 20º in the HOB angles there was a high accuracy for each subject and all classifiers. In particular, the accuracy was > 80% in classifying postures using the Naïve-Bayes classifier for all subjects. The corresponding accuracy using the KNN classifier resulted ≥ 90% in 8/10 subjects, with the remaining two subjects showing an accuracy value of 70% and 74%. SVM resulted in accuracy values of > 80% in 8/10 subjects with the remaining two resulting in an accuracy of 71% and 69%, respectively.

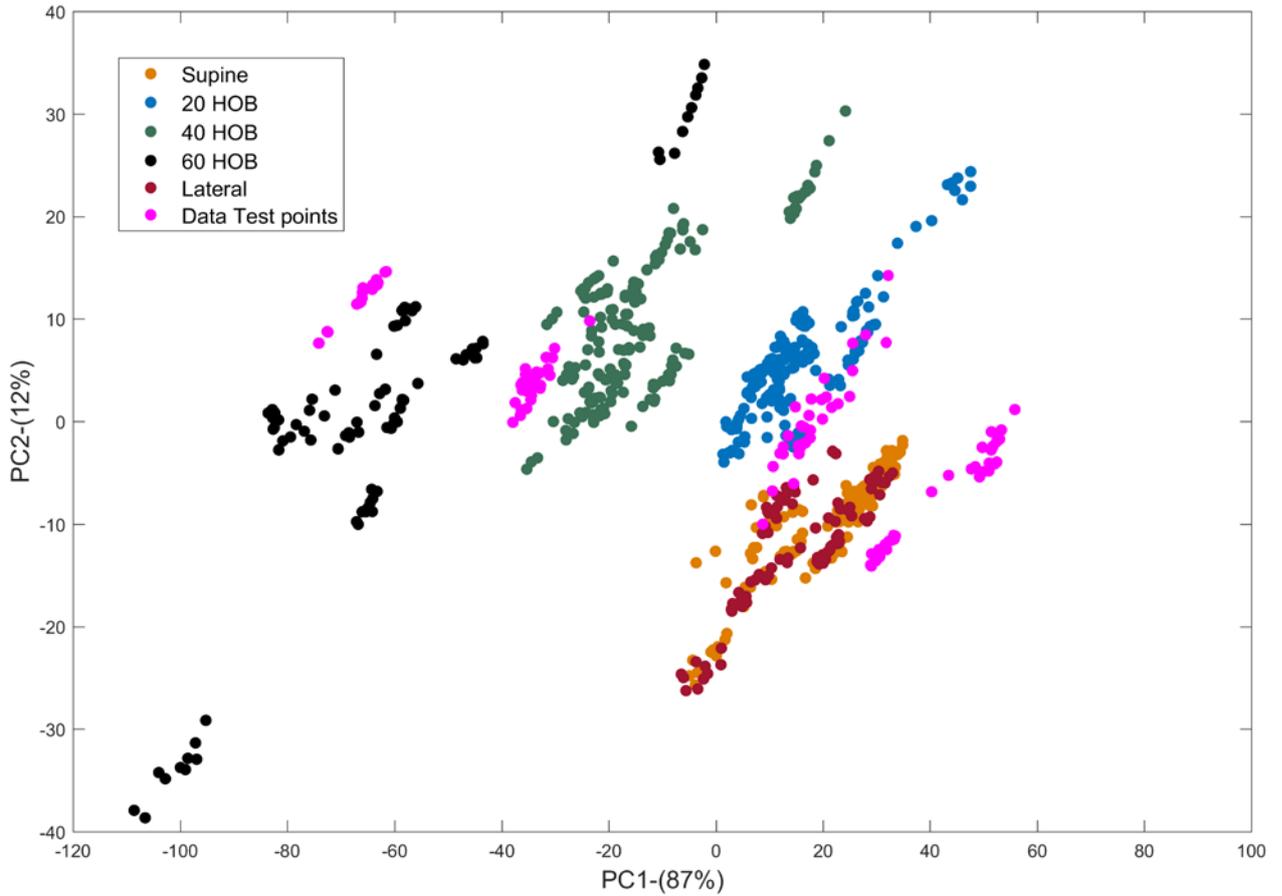

**Fig. 4:** Signals corresponding to the training data set (9 subjects) projected onto the first two principal components, PC1 and PC2, with their corresponding variance in brackets. Each posture is represented by a spatially distributed coloured cluster. Signals from one subject from the testing group (data points in pink) were projected onto the training PCs dimensional space. (For interpretation of the references to colour in this figure legend, the reader is referred to the web version of this article.)

**Table 1:** Percentage accuracy in classifying the range of postures for all classifiers.

| Subjects | Accuracy [%] | | |
|---|---|---|---|
| | **Naïve-Bayes** | **KNN** | **SVM** |
| **1** | 90 | 94 | 97 |
| **2** | 95 | 97 | 91 |
| **3** | 98 | 90 | 86 |
| **4** | 89 | 92 | 90 |
| **5** | 97 | 93 | 97 |
| **6** | 83 | 97 | 82 |
| **7** | 93 | 90 | 97 |
| **8** | 98 | 74 | 71 |
| **9** | 100 | 98 | 100 |
| **10** | 82 | 70 | 69 |

## 4. Discussion

This study has detailed the application of intelligent data processing of biomechanical signals depicting changes in lying posture from angles of body segments (actimetry) and pressures measured at the interface between the body and support surface i.e. contact area of pressures > 20mmHg. The derivative of signals was assessed to identify changes in posture in both sagittal and lateral planes. A series of machine learning algorithms in the form of Naïve-Bayes, KNN and SVM classifiers were applied to a set of data involving signals derived from an actimetry system and interface pressure distribution estimated from a high resolution sensing array. A cross-validation technique was applied using each machine learning algorithm, revealing that the training data could provide a robust means of classifying the data. Subsequently, an adapted protocol was used to provide test data, which was observed to correspond with the clusters derived from the training phase (Fig. 4). The resulting classification accuracy of the test data ranged between 82% 100%, 70%-98% and 69%-100% for the three classifiers, respectively (Table 1), with Naïve-Bayes classifier showing the highest accuracy in classifying the range of static postures. A value > 80% could represent a benchmark by which the majority of the postures can be monitored.

Findings revealed that the derivative of the signal representing the trunk tilt angles correctly identified the changes in posture for all subjects, as characterised by a transient increase in the magnitude of the derivative at each corresponding change in posture (Fig. 3A). By contrast, the derivative of the contact area signal generally revealed less distinct changes in magnitude when evaluating changes in posture involving HOB angles < 20º and thus it proved problematic in identifying their occurrence (Fig. 3B). Accordingly, the trunk tilt angles derivative, resulting from the sum of sagittal and lateral signal derivatives, was considered to represent a more robust means to automate the detection of the changes in posture in the test data. Indeed this approach, based on the derivative of biomechanical parameters, has been applied in several other areas of the biomedical field, for example, for the detection of the different gait phases (Taborri et al., 2016).

Previous research have utilised intelligent data processing from machine learning algorithms to classify a range of lying and sitting postures and their transitions from the distribution of pressure at the subject-support interface (Foubert et al., 2012; Kim et al., 2018; Matar et al., 2020; Rus et al., 2017; Wai et al., 2010; Yousefi et al., 2011; Zemp et al., 2016). These approaches adopted and the results reported are summarised in Table 2.

**Table 2:** Summary of relevant studies classifying lying and sitting postures.

| Study | Data used for classification | Classifier(s) | Accuracy | Changes in posture |
|---|---|---|---|---|
| Wai et al. (2010) - *Lying postures* | i) Raw pressure values<br>ii) Eigen vectors<br>iii) Mean, variance, standard deviation, root mean square estimated in 9 ROIs | SVM | > 70%<br>> 50%<br><br>> 60% | ✗ |
| Yousefi et al. (2011) - *Lying postures* | Binary pressure images projected in PCA space | KNN | > 97% | ✗ |
| Kim et al. (2018) - *Sitting postures* | Heat map of pressure distribution | Naïve - Bayes<br>SVM | > 85%<br>> 90% | ✗ |
| Duvall et al. (2019) - *Lying postures* | Weight measured by four cells placed under the legs of the bed | KNN | > 95% | ✓ |
| Zemp et al. (2016) - *Sitting postures* | Median of the force data divided by the subject's body weight and backrest angles | SVM | > 70% | ✗ |
| Foubert et al. (2012) - *Lying to sitting* | i) WNAS<br>ii) COP displacements | SVM and KNN | > 90%<br>> 75% | ✓ |
| Matar et al. (2020) - *Lying postures* | Oriented gradient and local binary patterns estimated from pressure distribution | Artificial neural network | > 97% | ✗ |
| Present study | Eigen vectors estimated from biomechanical parameters derived from actimetry systems and pressure distribution | Naïve-Bayes<br>KNN<br>SVM | > 80%<br>≥ 70%<br>≥ 69% | ✓ |

It is evident that the present findings are comparable with previous studies, with high accuracy values in postures classification reported for the detection of range of static postures. However, only two studies have detected the transition phases between static postures. Foubert et al. (2012) have used lateral and longitudinal displacements of the centre of pressure estimated from pressure distribution, reporting an accuracy of > 90%. By contrast, a separate study utilised the total weight on the bed measured by using a system involving four load cells (Duvall et al., 2019). Their results reported that a change of 7lb (3.2kg) in the measured weight within a temporal window of 7secs was able to detect the changes in posture with an accuracy of 98%. Furthermore, limitations of many previous studies included the short-term estimation of the pressure distribution (up to tens of seconds) and the limited range of supine postures (i.e. supine, prone, left and right turn). There are, however, some studies which have evaluated a range of postures involving the elevation of the HOB angle (Yousefi et al., 2011) and data derived from a longer period i.e. 5 min of pressure monitoring (Kim et al., 2018). In addition, Zemp et al. (2016) utilised a composite data set involving force values acquired at the support surface normalised to individual body weight and the corresponding backrest tilt angle estimated with actimetry positioned on the backrest, for the detection of sitting postures. However, both parameters were acquired at a single time point.

The present study has applied an automated method to identify the occurrence and magnitude of movements based on signal derivative and machine learning algorithms. This could be achieved using either actimetry or pressure parameters. To date, these temporal data are unknown in many care settings.

It is inevitable that the use of able- bodied cohorts in a lab-based study precludes generalising the present findings to individuals, from specific sub-populations, deemed to be at risk of developing pressure ulcers i.e. the elderly, spinal cord injured and those managed in intensive care units. The study protocol was also limited in selecting a pre-determined order of relatively small postural changes (20º HOB increments) maintained for a relative short period of 10-20 min. Thus, future studies should examine random postures involving different HOB increments and the side-lying lateral posture typically adopted in clinical settings. The current method would require an improvement in accuracy and validation to account for random postures on specialised mattresses used by patients in both acute and community clinical settings where the recommended frequency and magnitude of movements are not strictly followed (Defloor et al., 2005; Woodhouse et al., 2019). This would support clinicians when informing clinical decision-making.

Technology to monitor individuals could provide critical means to detect posture and mobility. However, it is clear that the emergence of digital health strategies will necessitate the use of robust monitoring tools. Accordingly, continuous pressure monitoring represents an important tool which when integrated with support and feedback technologies could promote PU prevention through self-management and targeted care interventions (Tung et al., 2015). This would result in more efficient practice and a personalised approach. It could also be integrated with risk assessment to create a more objective means of PU risk. In addition, machine learning applied to large data sets derived from these technologies could provide a robust means for translation into indicators of posture and mobility associated with both frequency and magnitude of postures.

## 5. Conclusion

The present study has defined a methodology for classifying static lying postures and identifying transitions in between different postures. The combination of biomechanical parameters acquired using pressure monitoring and actimetry technologies were combined using data reduction and machine learning approaches. The combination of monitoring technologies and advanced algorithms offers the potential to track posture and mobility in individuals at risk of pressure ulcers, informing personalised care strategies. Further research is needed to establish the accuracy of the posture prediction involving clinical data sets in sub-groups of patients at risk of pressure ulcers e.g. spinal cord injured.

**Acknowledgments**

The postgraduate researcher (SC) was supported by a UK Engineering and Physical Science Research Council Case award in association with Sumed International, who provided the pressure mapping system (ForeSite PT). SC was also supported by European Pressure Ulcer Advisory Panel (EPUAP) Research funding in collaboration with the University of Grenoble-Alpes to support the development of the present study. There were no conflicts of interest in this study.

Ethical approval

University of Southampton Ethics was granted for the study (Ref: 26379)